\documentclass[aps, prl, 10pt, superscriptaddress, twocolumn]{revtex4-1}

\usepackage[utf8x]{inputenc}
\usepackage{amsmath,amssymb}
\usepackage{graphicx}

\DeclareMathOperator{\re}{Re}
\DeclareMathOperator{\im}{Im}
\DeclareMathOperator{\tr}{tr}

\def\onlinecite{\cite}

\begin{document}

\title{Free fermions on a line: asymptotics of the entanglement entropy and 
entanglement spectrum from full counting statistics}

\author{Roman S\"usstrunk}
\affiliation{Institute for Theoretical Physics, ETH Zurich, 8093 Zurich, Switzerland}
\author{Dmitri A.~Ivanov}
\affiliation{Institute for Theoretical Physics, ETH Zurich, 8093 Zurich, Switzerland}
\affiliation{Institute for Theoretical Physics, University of Zurich, 8057 Zurich, Switzerland}


\begin{abstract}
We consider the entanglement entropy for a line segment in the system of noninteracting
one-dimensional fermions at zero temperature. In the limit of a large segment length $L$,
the leading asymptotic behavior of this  entropy is known to be logarithmic in $L$. We
study finite-size corrections to this asymptotic behavior.
Based on an earlier conjecture of the asymptotic expansion for full counting statistics
in the same system, we derive a full asymptotic expansion for the von Neumann entropy and
obtain first several corrections for the Rényi entropies. Our corrections for the Rényi
entropies reproduce earlier results. We also discuss the entanglement spectrum in this
problem in terms of single-particle occupation numbers.
\end{abstract}



\maketitle

\paragraph{1. Introduction.---}

Entanglement is one of the central concepts of modern quantum mechanics and
quantum information theory. It characterizes the amount of correlations
between parts of a quantum system. In recent years, a progress
has been achieved in studying entanglement for a variety of models,
with the most detailed results available for one-dimensional systems,
see e.g. the review \cite{latorre:09}.

Entanglement can be introduced in a particularly simple way in the
case of a many-body system in a pure state, e.g., in the zero-temperature
ground state, which will always be assumed in this paper. Let such a
system be divided into two subsystems $A$ and $B$. Then the entanglement
may be characterized by the properties of the reduced density matrix $\rho_A$
of the subsystem $A$, which is obtained by tracing out the remaining 
degrees of freedom
\begin{equation}
 \rho_A=\tr_B\rho
\end{equation}
(here $\rho$ denotes the density matrix of the pure state of the total system).
The (von Neumann) entanglement entropy is 
then defined as the von Neumann entropy of $\rho_A$,
\begin{equation}
 \mathcal{S}^{(A)}=-\tr\rho_A\ln\rho_A\,.
\end{equation} 
Though characterizing entanglement by a single number is appealing, it falls short 
in representing its full complexity. A more complete description of entanglement may 
be given by the set of Rényi entropies
\begin{equation}
 \mathcal{S}^{(A)}_{\alpha}=\frac{1}{1-\alpha}\ln\tr\rho_A^{\;\alpha}
 \, , \quad  \alpha \geq 0
 \, , \quad  \alpha \neq 1
\end{equation}
(the von Neumann entropy can then be expressed as the limit 
$\mathcal{S}^{(A)}=\lim_{\alpha\to 1} \mathcal{S}^{(A)}_{\alpha}$).
 
Since the total system is assumed to be in a pure state, these definitions can be shown
to be symmetric with respect to the interchange of the subsystems $A$ and $B$:
$\mathcal{S}^{(A)}=\mathcal{S}^{(B)}$ for both von Neumann and R\'enyi entropies \cite{latorre:09},
so we shall drop the superscript $(A)$ or $(B)$ in our notation below.

Equivalently, entanglement may be characterized by the spectrum of the reduced
density matrix $\rho_A$ (which coincides with the spectrum of $\rho_B$ for a pure state)
\cite{li:08:calabrese:08:fidkowski:10:pollmann:10}.
Like the full knowledge of the Rényi entropies, the entanglement spectrum allows to determine 
the state of the system up to unitary transformations in the subsystems $A$ and $B$. In this sense, 
the Rényi entropies and the entanglement spectrum contain the full information about entanglement.

The problem of calculating the entropies or the entanglement spectrum simplifies in the
case of {\em noninteracting particles} (bosons or fermions). In this case, the reduced 
density matrix ($\rho_A$ or $\rho_B$)
can be factorized into density matrices of individual single-particle levels\cite{d-matrix}, 
and both the entanglement spectrum and the entropies may be expressed in terms of {\em single-particle}
occupation numbers $p_i$. In the case of noninteracting {\em fermions}, the entropies are given
by
\begin{equation}
 \mathcal{S}=-\sum_i \left[ p_i \ln p_i + (1-p_i) \ln (1-p_i) \right]
\label{vN-noninteracting}
\end{equation} 
for the von Neumann entropy and
\begin{equation}
 \mathcal{S}_\alpha=\frac{1}{1-\alpha} \sum_i \ln \left[ p_i^\alpha + (1-p_i)^\alpha \right]
\label{Renyi-noninteracting}
\end{equation} 
for the Rényi entropies. The sums over $i$ can be converted into integrals over
$p_i$ [Eqs.\ (\ref{eq:VNInt}) and (\ref{eq:RenInt}) below] by introducing the 
spectral density of the occupation number
\begin{equation}
\mu(p) = \sum_i \delta(p - p_i)\, .
\end{equation}
This spectral density, together with the entanglement entropies 
(\ref{vN-noninteracting}) and (\ref{Renyi-noninteracting}),
in the model of noninteracting one-dimensional fermions,
will be the main object of our study.

Note that, in the case of noninteracting particles, the same spectrum of occupation 
numbers $p_i$ defines the full counting statistics (FCS) of the number of particles
in each of the two subsystems. This observation was used in 
Refs.~\onlinecite{song:11:12:klich:09:calabrese:12}
to establish an exact relation between the FCS and the entanglement spectrum. In the case
of noninteracting fermions, both the FCS and the entanglement spectrum can be expressed
in terms of the spectrum of a single-particle correlation matrix (in the context of
FCS, such a decomposition was done in Ref.~\onlinecite{abanov:08:09} on the basis of
the Levitov-Lesovik determinant formula \cite{levitov:93:96}).

Moreover, for noninteracting fermionic systems with translational invariance, the 
corresponding spectral problem involves matrices of Toeplitz type. Therefore, the
asymptotic behavior of FCS and entanglement spectrum in the limit of a large
subsystem size may be obtained with the help of the theory of Toeplitz determinants.
A prominent example is the spin-1/2 $XX$ chain \cite{jin:04,its:09}, 
which can be mapped to a system of noninteracting fermions via a Jordan-Wigner transformation.
In many interesting one-dimensional situations (including free fermions), the relevant
Toeplitz matrix has Fisher-Hartwig singularities, and the asymptotic behavior of its
determinant can be found using the Fisher-Hartwig conjecture \cite{basor:91:deift:11}.
While the leading asymptotic behavior of entanglement and FCS can be 
obtained by choosing the main Fisher-Hartwig branch, finding subleading contributions 
requires more work.  Recently, corrections to the entanglement entropies 
accurate to order $\mathcal{O}(L^{-3})$ (for a block of size $L$)
have been computed for the spin-1/2 $XX$ chain \cite{calabrese:10}
and in the continuous limit \cite{calabrese:11}.

Furthermore, in the continuous limit, 
a full asymptotic expansion
of the corresponding Toeplitz determinant was conjectured in Ref.~\onlinecite{ivanov:11} in the context of FCS.
Based on the matrix Riemann-Hilbert problem \cite{cheianov:04} and, independently, 
on the Painlevé V equation in the Jimbo-Miwa form \cite{jimbo:80:tracy:93},
an expansion was constructed for the FCS generating function of the particle number 
on a line interval for one-dimensional free fermions in the zero-temperature ground state.
Using the periodicity conjecture for the expansion (not proven, but verified up to high orders
in $1/L$), the asymptotic expansion was written in an explicitly periodic Fisher-Hartwig 
form \cite{ivanov:11}.
Instead of selecting the leading Fisher-Hartwig branch, all the branches were combined 
to obtain a full expansion to all orders in $1/L$, taking into account the switching of 
branches intrinsically.

We use the relation between FCS and entanglement entropies 
to carry over the full expansion conjectured in Ref.~\onlinecite{ivanov:11} 
of the FCS generating function to the problem of finding the entanglement 
entropies and the entanglement spectrum for free fermions on a line. 
In particular, we find the power-law asymptotic expansion for the von Neumann
entropy $\mathcal{S}$, compute first several coefficients, and present an algorithm for
calculating the coefficients to an arbitrary order. A similar approach to
the Rényi entropies $\mathcal{S}_\alpha$ produces an expansion 
with oscillating terms. For the Rényi entropies, we only compute the lowest-order terms,
which agree with the previously available results \cite{calabrese:10,calabrese:11}.
We also find finite-size corrections to the spectral density of single-particle 
occupation numbers $\mu(p)$.

The physical motivation for studying finite-size corrections to the entanglement
entropies is twofold. First, in critical one-dimensional systems, the form
of those corrections is related to the scaling dimensions of operators in 
the corresponding conformal field theory (CFT) \cite{cardy:10}. Second, knowing the structure
of finite-size corrections is helpful for extracting the central
charge of the CFT from numerical computations of the entropies \cite{finite-size-numerics}.

The remaining parts of the paper are structured as follows. The next section
embodies our main results. Then we review the
asymptotic expansion of the FCS for one-dimensional free fermions. 
Subsequently we present the calculations
of the spectral density $\mu(p)$ and of the von Neumann and Rényi entropies. 
Finally we conclude by a discussion of our results.
The appendix includes details of the analysis of oscillating terms 
in the asymptotic expansions of the entanglement 
entropies.

\begin{figure}[t]
\centerline{\includegraphics[width=0.4\textwidth]{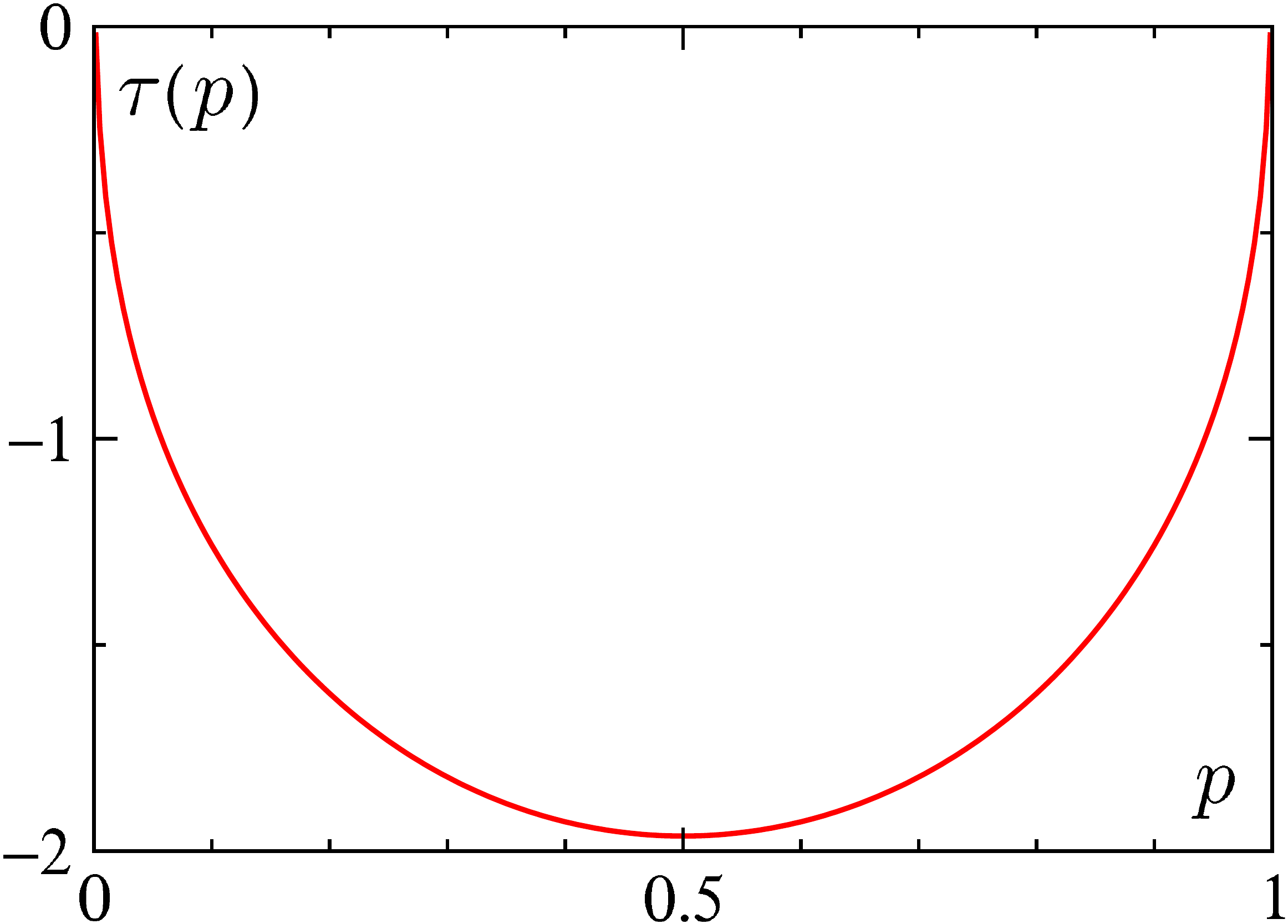}}
 \caption{The function $\tau(p)$ in Eq.~(\ref{DOS-result}). 
  At the end points of the interval $[0,1]$,
  $\tau(p)$ tends to infinity (logarithmically).}
 \label{fig:tau}
\end{figure}

\paragraph{2. Results.---}

Based on the conjecture for the FCS in Ref.~\onlinecite{ivanov:11}, 
we derive the asymptotic power series for the entanglement entropy of free fermions on a line
in the ground state:
\begin{equation}
 \mathcal{S}(x)=\frac{1}{3}\ln(2x)+\Upsilon + \sum_{n=1}^{\infty} s_{2n} x^{-2n}\, .
\label{eq:S-series}
\end{equation}
Here $x=k_F L$ ($L$ is the length of the line segment for which the entanglement is computed and
$k_F$ is the Fermi wavevector) and the constant $\Upsilon$ is given by Eq.~(\ref{eq:Upsilon}).
Note that this series contains only even powers of $1/x$. All the coefficients $s_{2n}$ are
rational numbers which can be computed to any given order in $n$ using the methods of
Ref.~\onlinecite{ivanov:11}. The first several coefficients are:
\begin{equation}
s_2=-\frac{1}{12}\, , \quad s_4=-\frac{31}{96}\, , \quad  s_6=-\frac{7057}{1440}\, .
\label{eq:S-coefficients}
\end{equation}
The leading term $(1/3)\ln(2x)$, the constant term $\Upsilon$, and the coefficient $s_2$ are known from earlier
works \cite{jin:04,calabrese:10,calabrese:11}.
In contrast to the Rényi entropies, there are no oscillating 
contributions to the von Neumann entropy at any order in $1/x$.

The calculation involves an expansion for the spectral density $\mu(p)$
based on the conjecture in Ref.~\onlinecite{ivanov:11}. Away from the end points $p=0$ and $p=1$
(a precise condition is formulated below),
the spectral density has a quasiclassical structure with locally nearly equidistant levels.
The smooth (nonoscillating) part of the spectral density is given by
\begin{equation}
 \bar\mu(p,x)=\frac{1}{\pi^2 \, p(1-p)}\left[\ln(2x)-\tau(p)\right] + \mathcal{O}(x^{-1})\, ,
\label{DOS-result}
\end{equation}
where
\begin{equation}
  \tau(p)=\varphi'\left[\frac{1}{2\pi}\ln\left(\frac{1-p}{p}\right)\right]\, ,
\label{eq:tau}
\end{equation}
\begin{equation}
 \varphi(\xi)=\arg\left[
 \Gamma \left(\frac{1}{2} + i\xi \right) \right]\, ,
\label{eq:varphi}
\end{equation}
and prime denotes the derivative of $\varphi(\xi)$ with respect to its argument.
The function $\tau(p)$ is plotted in Fig.~\ref{fig:tau}.

\paragraph{3. FCS of free one-dimensional fermions.---}

We consider free 
spinless fermions on a continuous line.
The temperature is assumed to be zero, i.e. the system is in the ground state 
characterized by the Fermi wavevector $k_F$. 
We will study the entanglement between two subsystems: 
an interval of length $L$ and the remainder of the line.
Both FCS and the entanglement in this setup depend only on the dimensionless parameter $x=k_FL$. For example,
the average number of particles on the line segment is given by $\left< N \right>=x/\pi$. The FCS 
generating function for the probability distribution of the particle number $N$,
\begin{equation}
 \chi(\kappa,x)=\left<e^{i\, (2\pi\kappa)\, N}\right>,
\end{equation}
was conjectured in Ref.~\onlinecite{ivanov:11} to be given by
\begin{equation}
 \chi(\kappa, x) = \sum_{j=-\infty}^{\infty} \chi_*(\kappa+j, x) \, , 
\label{eq:fcsExpansion1}
\end{equation}
\begin{multline}
 \chi_*(\kappa,x) =\exp\biggl[2i\kappa x - 2\kappa^2 \ln x + C(\kappa) \\
  +\sum_{n=1}^{\infty} f_n(\kappa)\, (ix)^{-n} \biggr]\, , 
  \label{eq:fcsExpansion2}
\end{multline}
\begin{equation}
 C(\kappa) =\ln \left[ G(1+\kappa)^2 G(1-\kappa)^2 \right] - 2\kappa^2 \ln 2\, , 
 \label{eq:defnC}
\end{equation}
where $G(z)$ denotes the Barnes G-function and $f_n(\kappa)$ are polynomials in $\kappa$, computable
order by order. For our purpose, we will use the logarithm of this expansion, which for $-1/2<\kappa <1/2$ takes the form
\begin{multline}
\ln\chi(\kappa,x)=2i\kappa x - 2\kappa^2 \ln x + C(\kappa) \\
+ \sum_{n=1}^\infty \; \sum_{m=-\left\lfloor \frac{n}{2}\right\rfloor}^{\left\lfloor \frac{n}{2}\right\rfloor} 
C_{n,m}(\kappa)\, x^{-n-4m\kappa}e^{2imx}\, ,  
 \label{eq:fcsLog}
\end{multline}
where $\left\lfloor \cdot \right\rfloor$ denotes the integer part of the argument. 

The coefficients $C_{n,m}(\kappa)$ can be expressed in terms of the polynomials $f_n(\kappa)$
order by order. They are also linearly related to the coefficients $R_{n,m}(\kappa)$ used
in Ref.~\onlinecite{ivanov:11} for the expansion of the derivative (in $x$) of 
Eq.~(\ref{eq:fcsLog}).
In particular, $C_{n,0}(\kappa)=-(1/n)\, R_{n+1,0}(\kappa)$.

\paragraph{4. Entanglement spectrum.---}

The spectral density $\mu(p)$ can be obtained from the jump of $\ln\chi(\kappa,x)$ across
the line $\kappa=\pm 1/2$ (see, e.g., Ref.~\onlinecite{abanov:11}):
\begin{equation}
\mu(p)=-\frac{1}{4\pi^2\, p(1-p)} \frac{\partial}{\partial \kappa} \ln \chi(\kappa,x) 
\biggr|_{\kappa=-(\frac{1}{2}-\varepsilon)-i\xi}^{\kappa=+(\frac{1}{2}-\varepsilon)-i\xi}\, ,
\label{eq:entSpecY}
\end{equation}
where $\varepsilon$ is an infinitesimally small positive parameter and we
introduced the parameterization
\begin{equation}
 \xi=\frac{1}{2\pi}\ln\left(\frac{1-p}{p}\right)\, .
\label{eq:changeOfVariables}
\end{equation}
Inserting Eq.~(\ref{eq:fcsLog}) into Eq.~(\ref{eq:entSpecY}) and
using the symmetry of the generating function $\chi(-\kappa,x) = \chi^* (\kappa,x)$,
we arrive at
\begin{multline}
 \mu(p) = - \frac{1}{2\pi^2\, p(1-p)} \re
\frac{\partial}{\partial \kappa}\Biggl[2i\kappa x - 2 \kappa^2\ln x + C(\kappa) \\
  +\sum_{n=1}^\infty \; 
\sum_{m=-\left\lfloor \frac{n}{2}\right\rfloor}^{\left\lfloor \frac{n}{2}\right\rfloor} 
C_{n,m}(\kappa)\, x^{-n-4m\kappa}e^{2imx} 
\Biggl]_{\kappa=\frac{1}{2}-i\xi}\, .
\label{eq:muExp}
\end{multline}

Note that the $x$ dependence of each term in Eq.~(\ref{eq:muExp}) is known. 
The coefficients at nonoscillating terms are determined by $C_{n,0}$, so that
the smooth (nonoscillating) part of $\mu(p)$ can be calculated as
\begin{multline}
\bar\mu(p) = \frac{1}{\pi^2\, p(1-p)}
\Big(\ln(2x) - \tau(p) \\
 + \sum_{n=1}^{\infty} \re C_{n,0}^\prime \left(\frac{1}{2}-i\xi\right)\, x^{-n} \Big) \, ,
\label{eq:mubar} 
\end{multline}
where $\tau(p)$ is given by Eqs.\ (\ref{eq:tau})--(\ref{eq:varphi})
and the prime denotes the derivative of $C_{n,0}(\kappa)$ with respect to $\kappa$.
The first two terms in this expansion give the announced result (\ref{DOS-result}).

Oscillating terms may, in turn, be collected by the ``diagonals''
$C_{2n,-n+l}(\kappa)$ and $C_{2n+1,-n+l}(\kappa)$ with $l=0,1,2,\dots$,
contributing terms of the orders $x^{-2l}$ and $x^{-2l-1}$, respectively.
The first two diagonals (with $l=0$) are easy to sum. By calculating the logarithm of the
series (\ref{eq:fcsExpansion1})--(\ref{eq:fcsExpansion2}) and using 
$f_1(\kappa)=2\kappa^3$ (see Ref.~\onlinecite{ivanov:11}), we find
\begin{align}
 C_{2n,-n}(\kappa) &=\frac{(-1)^{n+1}}{n}\, e^{n[C(\kappa-1) - C(\kappa)]}\, , \label{eq:diag1}\\
 C_{2n+1,-n}(\kappa) &= 2i (-1)^{n+1} (3\kappa^2 - 3\kappa + 1)\, e^{n[C(\kappa-1) - C(\kappa)]}\, .
\label{eq:diag2}
\end{align}
Adding those contributions converts the continuous spectrum (\ref{eq:mubar})
into a sum of delta functions. For example, taking into account the diagonals
(\ref{eq:diag1}) and (\ref{eq:diag2}) results in
\begin{equation}
\mu(p)=\sum_n \delta\left[\Phi(p,x) - \pi \left(n+\frac{1}{2}\right)\right]\, 
\frac{\partial\Phi}{\partial p} \, ,
\label{eq:muPhase}
\end{equation}
where
\begin{equation}
\Phi(p,x)=x + 2\xi\ln(2x) - 2\varphi(\xi) + \left(3\xi^2-\frac{1}{4}\right) x^{-1} 
+ \mathcal{O} \left(x^{-2}\right)
\label{eq:Phi}
\end{equation}
and $\varphi(\xi)$ is given by Eq.~(\ref{eq:varphi}).

Note that the spectrum (\ref{eq:muPhase}) has a quasiclassical nature:
the positions of quantum levels are determined by a 
quantization rule of Bohr-Sommerfeld type.
The resulting spectrum is regularly spaced with the average density
given by $(1/\pi)\, \partial\Phi/\partial p$. This can be explained by
the fact that the diagonals (\ref{eq:diag1}) and (\ref{eq:diag2}) stem,
in fact, only from the two leading Fisher-Hartwig branches in Eq.~(\ref{eq:fcsExpansion1}) 
at $\kappa=1/2$ (those with $j=0$ and $j=-1$). The spectrum is thus determined
from the condition that these two branches cancel each other, which naturally
leads to an expression of the form (\ref{eq:muPhase}). Including higher-order
Fisher-Hartwig branches produces modulations in the level spacing, but this
effect appears only at higher orders in $1/x$.

Note also that this expansion breaks down close to
the end points of the spectrum $p=1$ and $p=0$. In those regions $\tau(p)$ 
is large and therefore the density of states given by Eq.~(\ref{eq:mubar}) 
becomes formally negative: in fact, the expansion (\ref{eq:mubar}) is not applicable in those 
regions of $p$. Indeed, the expansion parameter in Eq.~(\ref{eq:mubar}) is $\xi/x$: this can
be seen from the (unproven) fact observed in Ref.~\onlinecite{ivanov:11} that
the polynomial $C_{n,0}$ has degree $n+2$ in $\kappa$ (and therefore in $\xi$).
Thus the expansion (\ref{eq:mubar}) is only applicable at $|\xi |\ll x$. 
Remarkably, this condition also guarantees the positivity of $\bar\mu(p)$.

Our results (\ref{DOS-result})--(\ref{eq:varphi}) and the quasiclassical
structure of the spectrum are consistent with the numerical studies of 
Refs.~\onlinecite{peschel:04:09}. In particular, the smooth part of 
the density of states
in the middle of the spectrum is
\begin{align}
& \bar\mu(p=1/2) = \frac{4}{\pi^2} \left( \ln x + b\right) + \mathcal{O} (x^{-1})\, , 
\notag \\
& b  = \ln2 - \varphi'(0) \approx 2.657\, ,
\end{align}
in agreement with the findings of those works.

\paragraph{5. Von Neumann and R\'enyi entropies.---}

Once the spectral density $\mu(p)$ is known, the entropies can be calculated using the integral forms of 
Eqs.\ (\ref{vN-noninteracting}) and (\ref{Renyi-noninteracting}):
\begin{equation}
 \mathcal{S}  =-\int_0^1 dp\, \mu(p)\, \left[ p\ln p+(1-p)\ln(1-p) \right] \label{eq:VNInt}
\end{equation}
for the von Neumann entropy and
\begin{equation}
 \mathcal{S_{\alpha}} =\frac{1}{1-\alpha}\int_0^1 dp\, \mu(p)\,
 \ln \left[ p^{\alpha}+(1-p)^{\alpha}\right] 
 \label{eq:RenInt}
\end{equation}
for the Rényi entropies.
Note that, even though the expansion (\ref{eq:muExp}) applies only at $|\xi|\ll x$,
we may integrate in Eqs.\ (\ref{eq:VNInt}) and (\ref{eq:RenInt}) 
from $\xi=-\infty$ to $\xi=+\infty$ (corresponding to $0<p<1$):
the contributions from large $\xi$ are exponentially smaller
than all the terms of the resulting series and may be neglected.

For the von Neumann entanglement entropy, oscillating contributions 
vanish at all orders in $1/x$ [the integral (\ref{eq:VNInt}) may
be closed in the upper or lower half plane of the variable $\xi$, 
see Appendix].
Only nonoscillating contributions survive and may be found  
by replacing $\mu(p)$ in the integral
(\ref{eq:VNInt}) by its nonoscillating part (\ref{eq:mubar}).
As a result, we find the power series
\begin{equation}
 \mathcal{S}(x)=\frac{1}{3}\ln(2x)+\Upsilon + \sum_{n=1}^{\infty} s_n x^{-n}\, ,
\label{eq:VNFullExpansion}
\end{equation}
where the coefficients are given by 
\begin{equation}
 s_n = \int_{-\infty}^{\infty} d\xi\, \frac{\pi \xi}{\cosh^2(\pi \xi)}\, 
\im C_{n,0} \left( \frac{1}{2} - i \xi \right)\, .
\end{equation}

The functions $C_{n,0}(\kappa)$ may be found from the results reported
in Ref.~\onlinecite{ivanov:11} or calculated order by order using the methods
developed in that work. In particular, it follows from the results of
Ref.~\onlinecite{ivanov:11} that $C_{2n+1,0}(\kappa)$ are polynomials odd 
in $\kappa$ with purely imaginary coefficients. Therefore, all the
odd terms in the expansion (\ref{eq:VNFullExpansion}) vanish, and we arrive
at the result (\ref{eq:S-series}).
Furthermore, since  $C_{n,0}(\kappa)$ are polynomials
with rational coefficients, all the coefficients $s_{2n}$ are
rational numbers. The first three nonzero coefficients can be obtained from
\begin{eqnarray}
  C_{2,0}(\kappa)&=& -\frac{5}{2} \kappa^4\, , \notag \\
  C_{4,0}(\kappa)&=& \frac{25}{16}\kappa^4+\frac{63}{4}\kappa^6\, ,\\
  C_{6,0}(\kappa)&=& -\frac{35}{8}\kappa^4-\frac{889}{12}\kappa^6-\frac{3129}{16}\kappa^8\, , \notag
\end{eqnarray}
which gives the result (\ref{eq:S-coefficients}).
Following this procedure [with $C_{2n,0}(\kappa)$ calculated using the method of
Ref.~\onlinecite{ivanov:11}], the coefficients $s_{2n}$ may be computed to
any order, one by one, in a straightforward way. 

The constant $\Upsilon$ is found to be
\begin{multline}
\Upsilon = -\frac{2}{\pi} \int_{-\infty}^{+\infty} d\xi \, \varphi'(\xi)
\left(\ln\left[2\cosh(\pi \xi)\right] - \pi \xi \tanh[\pi \xi] \right) \\
\approx 0.4950179081 \, ,
\label{eq:Upsilon}
\end{multline}
where the function $\varphi(\xi)$ is defined by Eq.~(\ref{eq:varphi}).
This expression for $\Upsilon$ can be shown to agree with
that found in Ref.~\onlinecite{jin:04}.

In contrast, for the Rényi entropies, the oscillating parts do not vanish and 
can be classified in terms of the poles of the integrand of Eq.~(\ref{eq:RenInt}),
see Appendix. 
The first orders [calculated using Eqs.~(\ref{eq:diag1}) and (\ref{eq:diag2})] are
given by Eqs.\ (\ref{eq:RenSeries}) and (\ref{eq:UpsilonAlpha}).
Calculating higher-order oscillating terms in the Rényi entropies would require
knowing higher order diagonals $C_{2n,-n+l}(\kappa)$ and $C_{2n+1,-n+l}(\kappa)$
with $l\ge 1$. Although each of those coefficient can be separately calculated 
(using the methods of Ref.~\onlinecite{ivanov:11}), deriving general formulas 
(valid for all $n$) is a tedious task, and we do not attempt it here.

\paragraph{6. Numercial illustration.---}

\begin{figure}[t]
\centerline{\includegraphics[width=0.45\textwidth]{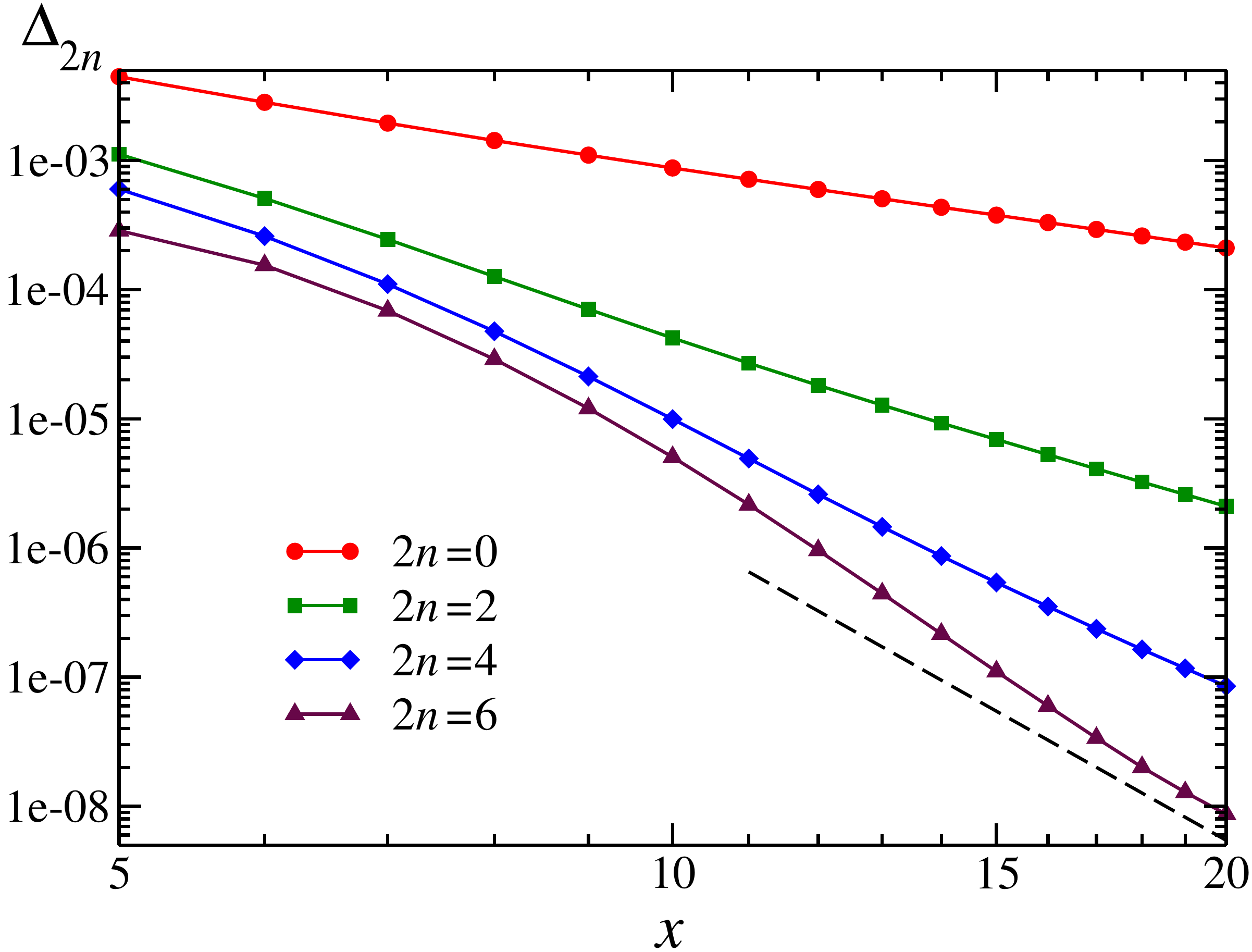}}
 \caption{The remainders of the asymptotic series (\ref{eq:S-series})
 as functions of $x$ (in the log-log scale). The dashed line indicates
 the slope of $x^{-8}$.}
 \label{fig:deviations}
\end{figure}

To illustrate our main result (\ref{eq:S-series})--(\ref{eq:S-coefficients}) and to 
perform an additional check of the expansion conjectured in Ref.~\onlinecite{ivanov:11},
we have also computed the von Neumann entropies $\mathcal{S}(x)$ numerically
and compared them to our analytical expansion (\ref{eq:S-series})--(\ref{eq:S-coefficients}). 
The numerical computation was performed in the lattice model
(considered, e.g., in Ref.~\onlinecite{abanov:11}) for blocks containing up to
1000 sites and then extrapolated to the continuous limit. This allowed us to calculate
$\mathcal{S}(x)$ for $x\in [5,20]$ with the error bars not exceeding $10^{-9}$. In 
Fig.~\ref{fig:deviations} we plot the remainder of the asymptotic series (\ref{eq:S-series})
$\Delta_{2n}=(1/3)\ln(2x)+\Upsilon + \sum_{m=1}^{n} s_{2m} x^{-2m} - \mathcal{S}(x)$
as a function of $x$. One can see that the remainders indeed decay
as powers of $x$: in particular, $\Delta_6$ decays as $x^{-8}$, in agreement with our
analytical prediction.

\paragraph{7. Summary and discussion.---}

In this paper, we have used the asymptotic expansion of the FCS generating function 
for a line segment of one-dimensional free fermions to determine the asymptotic expansion 
of the entanglement entropy and the entanglement spectrum in the same system. The main 
result is the asymptotic power series in $1/x$ for the von Neumann entropy. Our method 
also allows to construct finite-size corrections for the Rényi entropies 
(we only do it to the lowest order, where we reproduce the known results)
and gives an expansion for the spectrum of the single-particle correlation matrix.

Our results are based on the expansion conjectured (not rigorously proven)
in Ref.~\onlinecite{ivanov:11}, and therefore also have the status of conjecture. 
Two elements of the proof were missing in Ref.~\onlinecite{ivanov:11}. First,
the periodicity relations on the expansion coefficients [which allows to convert
the expansion into an explicitly periodic form (\ref{eq:fcsExpansion1})] were
not proven but only checked analytically up to the 15th order in $1/x$.
Second, the expansion (\ref{eq:fcsExpansion1})--(\ref{eq:fcsExpansion2})
was not extended to the line $\re(\kappa)=1/2$: the point where the switching
of the Fisher-Hartwig branches takes place and where we need the expansion
for calculating the entropies. An extension of the expansion to this line is
however a very plausible conjecture, since the expansion itself is
regular at this line;
it is also supported by a numerical study on the more general 
lattice model \cite{abanov:11} and by our numerical
computations of the von Neumann entropy (Fig.~\ref{fig:deviations}).
We thus conjecture that our results 
are in fact exact expressions for the model considered.

\begin{acknowledgments}
We thank P.~Calabrese, V.~Eisler, and I.~Peschel for helpful comments on the manuscript.
\end{acknowledgments}


\paragraph{Appendix: Oscillating contributions to the entropies.---}

In this appendix, we treat the oscillating (in $x$) terms in the
expansions of the von Neumann and Rényi entanglement entropies.
For the von Neumann entropy (\ref{eq:VNInt}), all the
oscillating terms vanish, provided the expansion conjectured
in Ref.~\onlinecite{ivanov:11} is correct. For the Rényi
entropies (\ref{eq:RenInt}), there are oscillating terms
decaying as $\alpha$-dependent powers of $x$.

Oscillating terms in the entropies are obtained by 
substituting the terms of the expansion (\ref{eq:muExp}) with
a given oscillation frequency $m$ into the
integrals (\ref{eq:VNInt}) and (\ref{eq:RenInt}).
The integrals are further calculated by using $\xi$
as the integration variable, integrating by parts
and closing the integration contour in the upper 
(lower) half plane for $m>0$ ($m<0$, respectively).

In the case of the von Neumann entropy, this
produces terms of the form
\begin{multline}
\im\Big[ e^{2imx} x^{-n-2m} \int_{-\infty}^{\infty} d\xi \,
\frac{\pi \xi}{\cosh^2 (\pi \xi)} \\
\times C_{n,m} \left(\frac{1}{2} - i \xi \right)
e^{4 i m \xi \ln x} \Big]\, .
\label{eq:oscIntVN}
\end{multline}
Now the crucial ingredient of our discussion is the structure of the
coefficients $C_{n,m}(\kappa)$. It can be seen from the
explicit calculation in Ref.~\onlinecite{ivanov:11}
(using the Riemann-Hilbert method) that these coefficients
have the following form (assuming $m>0$):
\begin{eqnarray}
C_{n,m}(\kappa) & = & \tilde{c}_{n,m}(\kappa) e^{m[C(\kappa+1) - C(\kappa)]} \, , \\
C_{n,-m}(\kappa) & = & \tilde{c}_{n,-m}(\kappa) e^{m[C(\kappa-1) - C(\kappa)]} \, ,
\end{eqnarray}
where $C(\kappa)$ is defined in Eq.~(\ref{eq:defnC}) and $\tilde{c}_{n,m}(\kappa)$
are some polynomials in $\kappa$.

From this property, it follows that, at $m>0$, the coefficient $C_{n,m}(1/2 - i \xi)$ has zeroes
of degree $2m$ at all points $\xi= i(1/2 + r)$ for $r=0, 1, \ldots$, which compensate
the poles of degree two of the factor $\cosh^{-2} (\pi \xi)$ in the integral (\ref{eq:oscIntVN}).
Therefore the integrand is analytic in the upper half plane where the contour
is closed, and the integral vanishes. Similarly, at $m<0$, the coefficient $C_{n,m}(1/2 - i \xi)$ 
has zeroes of degree $2m$ at all points $\xi= - i (1/2 + r)$ for $r=0, 1, \ldots$,
the integrand is analytic in the lower half plane, and the integral vanishes again.
We therefore conclude that the asymptotic expansion of the von Neumann entanglement entropy
has the form of a power series in $1/x$ (apart from the leading logarithm), without any
oscillating terms.

In the case of the Rényi entropies, the oscillating terms have the form
\begin{multline}
\im\Big[ e^{2imx} x^{-n-2m} \int_{-\infty}^{\infty} d\xi\,
\frac{\alpha \left[ \tanh (\pi \xi) - \tanh (\alpha \pi \xi) \right]}{1-\alpha} \\
\times C_{n,m} \left(\frac{1}{2} - i \xi \right) e^{4 i m \xi \ln x} \Big ]\, .
\label{eq:oscIntRen}
\end{multline} 
They contain additional poles at $\xi=\pm (i/\alpha)(1/2+n)$. These poles are
not compensated by zeroes of $C_{n,m}(1/2-i \xi)$ and produce oscillating
contributions to the entropy decaying as fractional ($\alpha$-dependent)
powers of $x$. A calculation of the first few terms
[based on the explicit expressions (\ref{eq:diag1}) and (\ref{eq:diag2})]
produces the result
\begin{multline}
 \mathcal{S}_{\alpha}(x) =\frac{1}{6}\left[1+\frac{1}{\alpha}\right]\ln(2x)+\Upsilon_\alpha +\frac{(\alpha+1)(3\alpha^2-7)}{96\alpha^3}x^{-2} \\
  +\sum_{n,j=1}^\infty\frac{(-1)^n}{\alpha-1}\,(2x)^{-\frac{2n(2j-1)}{\alpha}}
\left[\frac{\Gamma\left(\frac{1}{2}+\frac{2j-1}{2\alpha}\right)}{\Gamma\left(\frac{1}{2}-
\frac{2j-1}{2\alpha}\right)}\right]^{2n} \\
  \times \left[\frac{2}{n}\cos(2nx)+x^{-1}\left[1+3\frac{(2j-1)^2}{\alpha^2}\right]\sin(2nx)\right] \\
  + o (x^{-2})\, ,
\label{eq:RenSeries}
\end{multline}
where
\begin{multline}
\Upsilon_\alpha = -\frac{2}{\pi} \int_{-\infty}^{+\infty} d\xi\, \varphi'(\xi) \\
\times \frac{\ln\left[2\cosh(\pi \xi \alpha)\right] - \alpha \ln [ 2 \cosh(\pi \xi) ]}{1-\alpha} \, .
\label{eq:UpsilonAlpha}
\end{multline}
These corrections reproduce the results
of Ref.~\onlinecite{calabrese:10} (in the corresponding
continuous limit of the spin-1/2 $XX$ chain),
\onlinecite{calabrese:11}, and \onlinecite{jin:04}.


\end{document}